\documentclass[fleqn,usenatbib]{mnras}

\usepackage{newtxtext,newtxmath}

\usepackage[T1]{fontenc}

\DeclareRobustCommand{\VAN}[3]{#2}
\let\VANthebibliography\thebibliography
\def\thebibliography{\DeclareRobustCommand{\VAN}[3]{##3}\VANthebibliography}

\usepackage{graphicx}	
\usepackage{amsmath}	

\usepackage{xcolor}
\usepackage{soul}

\setstcolor{red}

\soulregister\citep7
\soulregister\citet7

\usepackage{natbib}

\title[A New Galactic CBSS Candidate]{A New Galactic Close Binary Supersoft Source Candidate}

\author[G. R. C. Barreteau et al.]{
G. R. C. Barreteau,$^{1}$\thanks{E-mail: gaelromain@usp.br}
M. P. Diaz,$^{1}$\thanks{E-mail: marcos.diaz@iag.usp.br}
P. E. Stecchini,$^{2}$
N. Palivanas,$^{3}$
A. S. Oliveira,$^{3}$
L. Mazzero,$^{2}$
\newauthor F. D'Amico,$^{2}$
and F. Jablonski$^{2}$
\\
$^{1}$Instituto de Astronomia, Geofísica e Ciências Atmosféricas, Universidade de São Paulo, Rua do Matão 1226, SP-05508-090, São Paulo, Brazil\\
$^{2}$Divisão de Astrofísica, Instituto Nacional de Pesquisas Espaciais, Av. dos Astronautas 1758, SP-12227-010, S.J. Campos, Brazil\\
$^{3}$Instituto de Pesquisa \& Desenvolvimento, Universidade do Vale do Paraíba, Av. Shishima Hifumi 2911, SP-12244-000, S.J. Campos, Brazil
}

\date{Accepted XXX. Received YYY; in original form ZZZ}

\pubyear{\the\year{}}

\begin{document}
\label{firstpage}
\pagerange{\pageref{firstpage}--\pageref{lastpage}}
\maketitle

\begin{abstract}
The identification of a new galactic Close Binary Supersoft Source (CBSS) candidate, whose main ID is ZTF\,J043609.49+364057.7, is presented on the basis of the analysis of its optical spectrum, light curve and Spectral Energy Distribution (SED). During a survey designed to find candidates to the aforementioned class independently of X-ray emissions, this object's optical spectrum was found to exhibit the typical emission features expected for a CBSS. The available photometric data show a quasi-sinusoidal modulation, with a period of  5.42\,h, compatible with the orbital modulation of the system. Although there is no current identification of X-ray emission from this source, its SED displays a highly absorbed bright \textit{FUV} component, given the estimated $E(B - V) \sim 0.7$\,mag. 
\end{abstract}


\section{Introduction}

The Close Binary Supersoft Sources (CBSSs) are a subset of the so-called Luminous Supersoft X-ray Sources (SSSs), a class mainly characterized by their intense emission in soft X-rays and extreme ultraviolet (\textit{EUV}), accompanied by very high bolometric luminosities of $ 10^{36}$ to $10^{38}$\,erg/s \citep{1997ARA&A..35...69K}. The SSS class was defined originally from the Roentgen Satellite (\textit{ROSAT}) observations of the Magellanic Clouds and M31, with CAL\,83 and CAL\,87 becoming the class archetypes. The observed X-ray fluxes of such objects were shown to be best described by a massive white dwarf (WD) accreting hydrogen-rich material at rates of $\sim 10^{-7} M_{\odot}$\,yr$^{-1}$, allowing for thermonuclear burning to take place on its surface in a semi-stable state \citep{1992A&A...262...97V}. The CBSS scenario is plausible for such process to occur; these systems are composed by a WD in a close orbit, have periods ranging from 5 hours to a few days, and a Main Sequence (MS) or post-MS companion star that fills its Roche-Lobe, resulting in a semi-detached binary that, eventually, provides the adequate accretion rate. The properties, variability and eruptive behaviour of these systems are still a matter of debate, as pointed out by \citet{2017JAFS....3...30B}, especially due to the evident heterogeneity among the studied members of the class. It is worth noting that the SSSs are expected to be type Ia Supernovae (SNe\,Ia) progenitors, thanks to the lack of nova outburst episodes and the probable increase of the WD's mass over time \citep{1996sao..reptR....D}.

In CBSSs, the X-rays produced on the WD's photosphere are reprocessed both by the geometrically (and optically) thick accretion disc \citep{1996LNP...472...65P}, and by the irradiated face of the tidally-locked companion, which is considerably heated. Therefore, spectral emission lines are expected to originate from the aforementioned regions, and in a radiative wind above the WD's photosphere (e.g. CAL 87, \citealp{1997ARA&A..35...69K}). Beyond the X-rays, CBSSs are also characterized by their optical spectra, usually displaying very intense H and He\,\textsc{ii} emission lines accompanied by the conspicuous high-ionization emission lines of N\,\textsc{v} and O\,\textsc{vi}. Notably, it is usual for their spectra to present an equivalent width (EW) ratio He\,\textsc{ii} $\lambda 4686$\,\AA\,/\,H\,$\beta$ $>$ 2, in contrast with other emission line spectra seen in most Wolf-Rayet (WR) stars and Magnetic Cataclysmic Variables (MCVs).

\begin{figure*}
    \centering
    \includegraphics[width=1\linewidth]{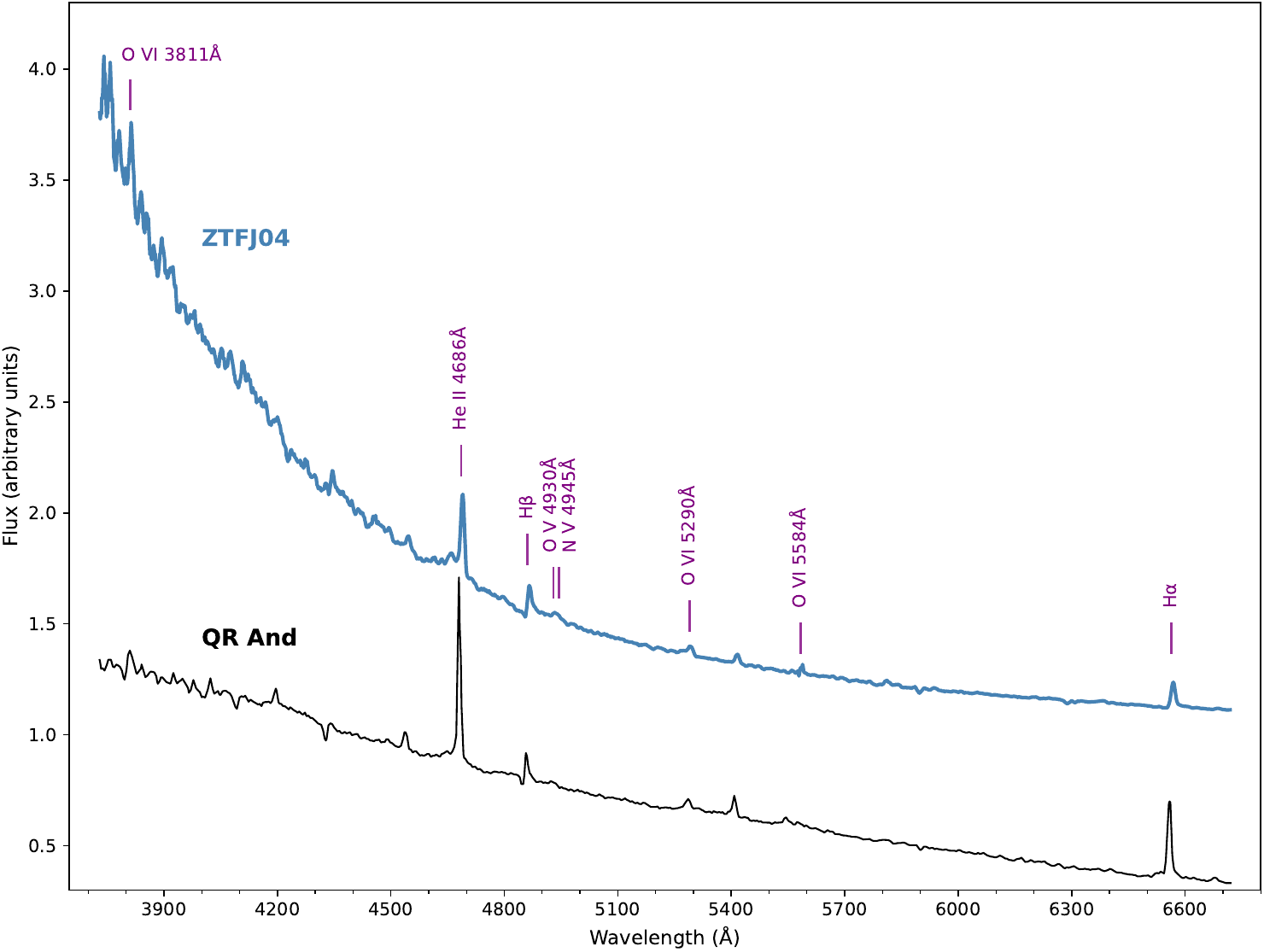}
    \caption{Optical spectrum of ZTFJ04, retrieved from LAMOST DR10, compared to QR\,And's spectrum from \citet{1995A&A...294L...1B}. Both were dereddened and normalized at 4400\,\AA. An offset of +1 was applied to ZTFJ04.}
    \label{fig:spec}
\end{figure*}

Currently only one non-Nova CBSS is known in the Milky Way; QR\,And \citep{1995A&A...294L...1B}. However, the CBSS model has also been proposed for other four galactic close binaries, despite certain heterogeneity amongst them, namely WX\,Cen \citep{1995AJ....110.1816D}, V\,Sge \citep{1999PASP..111...76D}, V617\,Sgr \citep{1999A&A...351.1021S}, and QU\,Car \citep{2008AJ....135.1649K,2014MNRAS.444.2692O}. One \textit{ROSAT} observation of V\,Sge in a low optical state showed a very soft source \citep{1998A&A...339L..21G}, although the luminosity could not be well determined. The other three objects have also been observed in X-rays and a SSS state has not been discovered yet, so it is possible that such a state is transient and occurs only rarely. These objects were also named as `V\,Sge stars', grouped together especially for their similar optical spectrum features and photometric variability \citep{1998PASP..110..276S}. They are considered galactic counterparts of the CBSS extragalactic sample, not necessarily identified by X-ray emission. These rare systems represent roughly half of all the known galactic SSSs. Meanwhile, there were 7 CBSSs identified in the Magellanic Clouds among 12 SSSs \citep{2000NewA....5..137G}, and there is a total of 89 SSSs identified in M31 \citep{2010ApJ...717..739O}, from which only a few are pointed to have some characteristics of the CBSS scenario. A possible explanation for the apparent discrepancy between the galactic and extragalactic samples is based on interstellar absorption, which plays a much stronger role in the attenuation of soft X-rays for most sources lying in the galactic disc than for targets in the mentioned nearby galaxies. Due to the reduced sample of known galactic CBSSs, the properties of this population and their role in the single-degenerate channel to SNe\,Ia are highly uncertain.

Given the possibility of a hidden population of galactic CBSSs, a survey based on their optical spectra is being conducted to search for these objects. By measuring He\,\textsc{ii}, O\,\textsc{vi}, and N\,\textsc{v} line fluxes relative to H\,$\beta$, a relatively small sample may be selected for visual inspection. Approximately half a million spectra from the Sloan Digital Sky Survey (SDSS; \citealp{2000AJ....120.1579Y}) and the Large Sky Area Multi-Object Fiber Spectroscopic Telescope (LAMOST; \citealp{2012RAA....12.1197C,2012RAA....12..723Z}) were analysed so far. Next, the first confirmed CBSS candidate from this survey, whose main ID is ZTF\,J043609.49+364057.7 (hereafter referred to as ZTFJ04) is described and contextualized.

\section{ZTF J043609.49+364057.7}

\begin{figure*}
    \centering
    \includegraphics[width=1\linewidth]{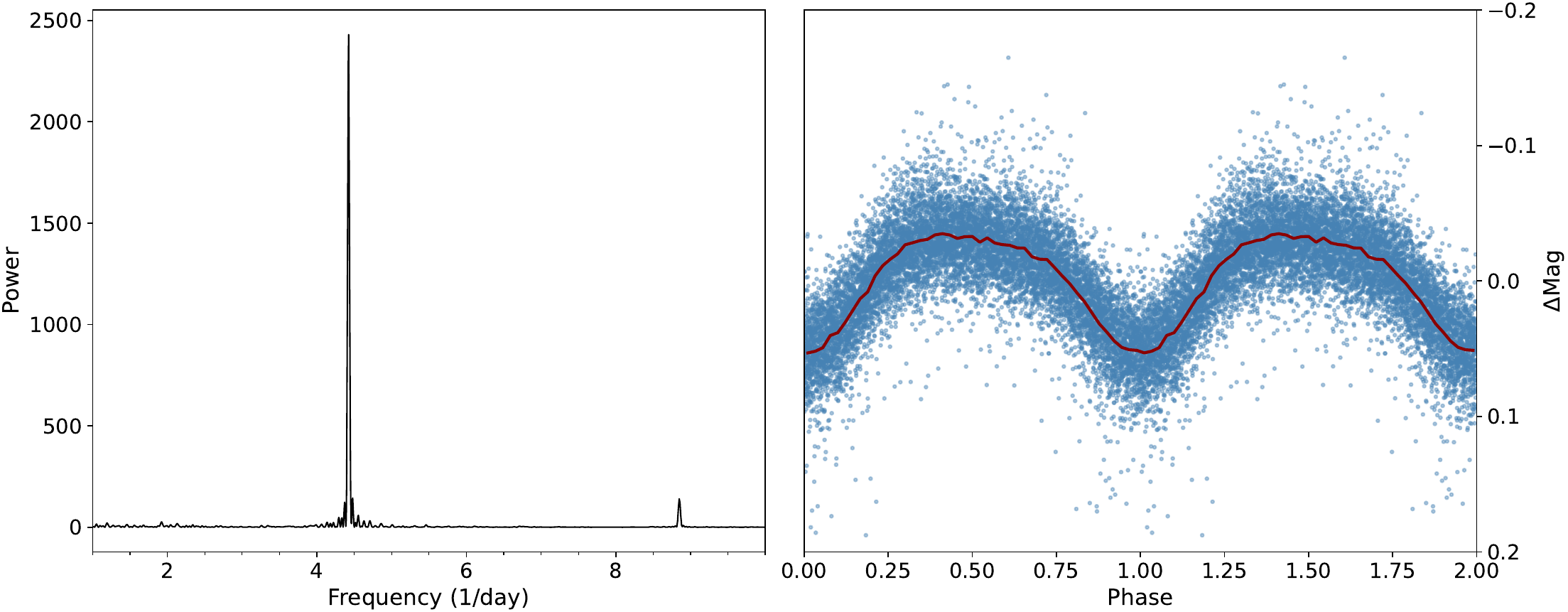}
    \caption{Left: Power spectrum of TESS sector 59 data. Right: Phase-folded TESS light curve obtained for the 5.424(19)\,h period. The red line is a Kalman-filtered average. Phase 0.0 is defined at the photometric minimum epoch BTJD $\equiv$ BJD-2457000.0 = 2910.488(7).}
    \label{fig:LC}
\end{figure*}

The optical spectrum of ZTFJ04 retrieved from LAMOST DR10 is shown in Figure \ref{fig:spec} in comparison to that of the well-established galactic CBSS QR\,And, obtained by \citet{1995A&A...294L...1B}. The spectra are corrected for an interstellar extinction of $E(B - V) \sim 0.7$\,mag, estimated for ZTFJ04 considering results from \citet{1992A&A...258..104A}, Bayestar 3D Dust Mapping \citep{2019ApJ...887...93G} and IRSA's Galactic Dust Reddening and Extinction Service \citep{2011ApJ...737..103S}, and $E(B - V) = 0.1$\,mag for QR\,And \citep{1995A&A...294L...1B}. Both share several similarities common to CBSS spectra. The Balmer and He\,\textsc{ii} lines are seen in strong emission, with EW($\lambda$4686\,\AA) $\simeq$ -6.1\,\AA \ and EW(H\,$\beta$) $\simeq$ -2.4\,\AA \ for ZTFJ04. The high-ionization emission lines of O\,\textsc{vi} ($\lambda\lambda$3811-34,5290,5584\,\AA), N\,\textsc{v} ($\lambda$4945\,\AA) and O\,\textsc{v} ($\lambda$4940\,\AA) are clearly visible. P\,Cyg profiles are present in the H\,$\beta$ lines, with close main velocities ($\sim - 670$\,km\,s$^{-1}$ for ZTFJ04 and $\sim - 630$\,km\,s$^{-1}$ for QR\,And). Despite the similarities, ZTFJ04's UV continuum is much steeper, a feature that could indicate that this system is softer than QR And. The lower He\,\textsc{ii} $\lambda$4686\,\AA \ to H\,$\beta$ ratio of ZTFJ04 may also corroborate this hypothesis, if the highly absorbed central source continuum peaks below 54\,eV and recombination emission prevails. Furthermore, \textit{Gaia} DR3 data \citep{2016A&A...595A...1G,2023A&A...674A...1G} provides a distance of 5.8$^{+1.9}_{-1.1}$\,kpc toward the galactic anti-centre, at low galactic latitude ($b = -7.2^o$), for ZTFJ04. It is noteworthy that this is an intrinsically bright system, with $M_V \simeq -0.21$, another close resemblance to QR\,And ($M_V \simeq -0.26$). 

The Transiting Exoplanet Survey Satellite (TESS, \citealp{2015JATIS...1a4003R}) observed the ZTFJ04 field in sector 59, in November and December, 2022. The data downloaded from the Mikulski Archive for Space Telescopes  (MAST) were used to construct the light curve (LC) from the Full-Frame images (FFI), using the Python package Lightkurve \citep{lightkurve} to perform aperture photometry. The most probable period found in the power spectrum, shown in the left panel of Figure \ref{fig:LC}, for the complete TESS series is 5.424(19)\,h. The corresponding phase-folded LC, displayed in the right panel of Figure \ref{fig:LC}, shows a quasi-sinusoidal modulation with an amplitude of $\sim 0.1$\,mag. Considering the power spectrum peak amplitude, we suggest that this is the correct choice for the orbital period. A slightly faster rise to maximum can be seen in the data at hand, resembling the LC shape of other CBSSs \citep{1999A&A...351.1021S,2023MNRAS.522.2732Z,2026JHEAp..5000467S}, although ZTFJ04 does not present a secondary eclipse as seen in these examples. Despite this small asymmetry, the LC shape is consistent with what would be expected from a significant contribution of an illuminated companion, with its modulated aspect along the orbital cycle. The phase-folded ($g-r$) colour index derived from the available Zwicky Transient Facility (ZTF) \citep{2019PASP..131a8003M} data does not show any meaningful pattern throughout the orbital period. 

Prior to this work, ZTFJ04 had already received two distinct classifications. It was identified as a variable star by the ZTF through optical photometry, being classified in its Catalog of Periodic Variable Stars as an eclipsing contact binary (W\,UMa type) with a 10.84\,h orbital period \citep{2020ApJS..249...18C}. However, it is clear from the analysis of the combined data presented that ZTFJ04 should not be a W\,UMa system. Its high luminosity ($M_V \simeq -0.21$) and strong high-ionization emission lines are incompatible with what would be expected for such a contact binary ($M_V=$ +2 to +6;  \citealp{Rucinski&Duerbeck}). It should be noted that the W\,UMa classification was given based solely on the automated analysis of the light curve built from low-cadence data. In turn, the analysis of much better sampled data from TESS reveals that the previous 10.84\,h period found is a multiple of the most probable 5.42\,h orbital period.

ZTFJ04's optical spectrum was labelled as an MCV by machine-learning filtering and visual inspection \citep{2020AJ....159...43H,2021ApJS..257...65S}. Later, the same method recognized it as a regular Cataclysmic Variable (CV) \citep{2023AJ....165..148H}. Although the high He\,\textsc{ii} to H\,$\beta$ ratio was measured, no comment was made on how unusual such ratio is for a CV, or even for an MCV. However, this is a common feature of CBSS spectra, together with the distinctive presence of O\,\textsc{vi} and N\,\textsc{v} emission lines, as previously pointed out. In addition, even the brightest MCVs are much less luminous than ZTFJ04.

\begin{figure}
    \centering
    \includegraphics[width=1\linewidth]{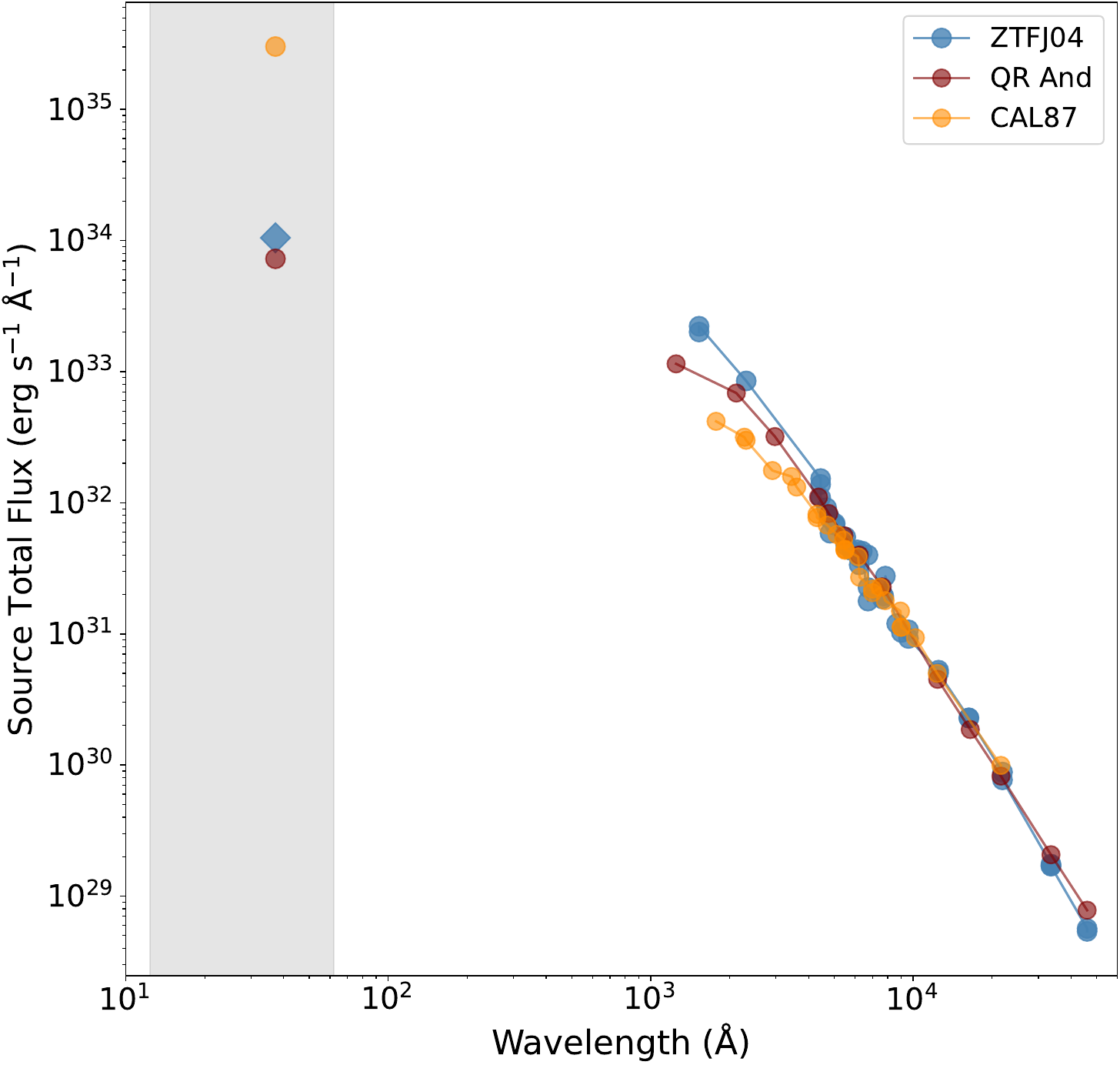}
    \caption{SED of ZTFJ04 compared with known CBSSs, corrected for extinction. The leftmost point of ZTFJ04 corresponds to a rough estimate of its soft X-ray luminosity, as described in text. The gray region represents the X-ray band over which the respective points are averaged (0.2 -- 1.0\,keV).}
    \label{fig:sed}
\end{figure}

The Spectral Energy Distribution (SED) of ZTFJ04 presented in Figure \ref{fig:sed} includes all the available data (which have attributed uncertainties) found in the public Strasbourg astronomical Data Center (CDS) SED tool (GALEX, POSS, PAN-STARRS, SDSS, \textit{Gaia}, WISE, APASS, GSC, FONAC, TESS \& 2MASS). Despite the lack of any detection in X-rays, there is a very luminous \textit{FUV} component, which is highly reddened. The strong resemblance between the optical spectra of ZTFJ04 and QR\,And in Figure \ref{fig:spec} suggests that both systems may have roughly similar X-ray to optical luminosity ratios. Using this comparison hypothesis and QR\,And's X-ray to optical ratio ($L_{\rm X}/L_{\rm opt}$) derived from the SED by \citet{https://doi.org/10.1002/asna.70147}, one may roughly estimate ZTFJ04's X-ray monochromatic luminosity, which is shown as an indicative point alongside the SED in Figure \ref{fig:sed}. The SEDs of QR\,And \citep{https://doi.org/10.1002/asna.70147} and CAL\,87 \citep{2026JHEAp..5000467S} are also presented for comparison.

A search for archival X-ray coverage of ZTFJ04 found three \textit{ROSAT}/PSPC All-Sky Survey observations covering the source position, with no detection at the optical coordinates. Count-rate upper limits were estimated from the PSPC images and exposure maps using a 3\,arcmin source aperture and a concentric 4\,--\,8\,arcmin background annulus. The source-aperture counts are consistent with the expected local background in all three observations; combining them gives a 90\% Poisson upper limit of 8.7\,$\times$\,10$^{-3}$\,ct\,s$^{-1}$.

This combined count-rate limit was converted with \textsc{xspec} \citep{xspec}, using a \textit{ROSAT}/PSPC-C response and absorbed blackbody models with $kT$\,=\,20\,--\,70\,eV and $N_{\rm H}$\,=\,3.0\,--\,6.2\,$\times$\,10$^{21}$\,cm$^{-2}$, into unabsorbed luminosity limits in the 0.2\,--\,1.0\,keV band, assuming a distance of 5.8\,kpc. This $N_{\rm H}$ range spans the Galactic H\,\textsc{i} value inferred from the optical reddening. The $E(B - V)$ versus hydrogen column density $N_{\rm H}$ relation for the galactic disc by \citet{Shull_etal2021} indicates  $N_{\rm H}\approx 4.3\times10^{21}$\,cm$^{-2}$. The resulting limits are shown in Figure~\ref{fig:upperlim}. For comparison, the 0.2\,--\,1.0\,keV luminosities expected from Eddington-limited blackbodies for 1\,M$_\odot$ WDs with solar abundances are indicated for $kT$ in the 20\,--\,70\,eV range. Also shown is the 0.2\,--\,1.0\,keV luminosity of ZTFJ04, gauged by assuming the $L_{\rm X}/L_{\rm opt}$ ratio of QR\,And.

\section{Discussion}

\begin{figure}
    \centering
    \includegraphics[width=1\linewidth]{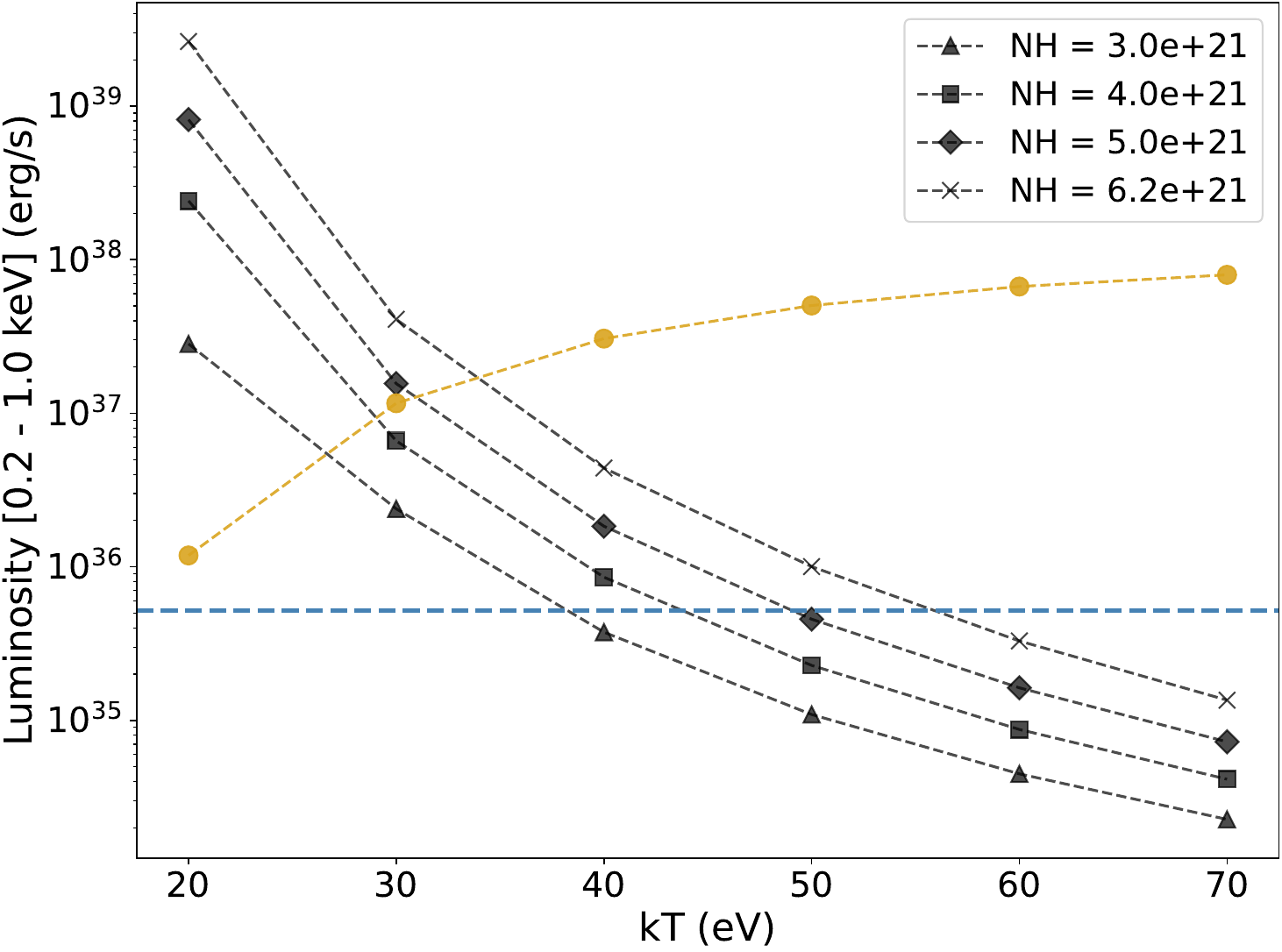}
    \caption{Upper limits for ZTFJ04's X-ray emission based on the non-detection from \textit{ROSAT} observations of the field, for different values of $N_{\rm H}$. The yellow curve indicates the luminosities in the 0.2\,--\,1.0\,keV band of a 1\,M$_\odot$ WD radiating at Eddington luminosity with the respective blackbody temperatures. The blue horizontal dashed line represents a possible estimate of ZTFJ04's X-ray luminosity in the referred band (see text).}
    \label{fig:upperlim}
\end{figure}

The spectrum similarity with QR\,And, known to be an extremely soft CBSS (with blackbody temperature of 20\,--\,22\,eV, as pointed by \citet{1995A&A...294L...1B} and \citet{2006ESASP.604..285R}), may suggest a low-temperature position for ZTFJ04 in Figure \ref{fig:upperlim}. Considering the Eddington luminosity curve and source upper limits, it is possible for ZTFJ04 to have a typical sub-Eddington CBSS luminosity and have remained undetected due to interstellar absorption. If the $L_{\rm X}/L_{\rm opt}$ ratio of ZTFJ04 is comparable to QR\,And one may find an upper limit for the temperature of the soft X-ray component $\lesssim$ 40\,--\,60\,eV, depending on absorption. On the other hand, the \textit{ROSAT} limits exclude the emission from a hot and/or luminous supersoft component. Besides the close similarity with the QR\,And spectrum, the high He\,\textsc{ii} $\lambda 4686$\,\AA\,/\,H\,$\beta$ ratio as well as the N and O ionization states are also seen in CAL\,83, RX\,J0513.9-6951, V\,Sge, WX\,Cen and V617\,Sgr, which require an ionizing radiation field with photon energy above 114\,eV.

Early estimates pointed to the probable existence of 100 to 1000 SSSs in the Galaxy \citep{1994ApJ...426..692R}. While a significant part of these may fall into the CBSS scenario, only five of such sources are currently proposed as galactic CBSSs, even after several X-ray surveys with sensitivity at the CBSSs' high-energy tail. Population synthesis of CBSSs in late-type galaxies by \citet{Chen_etal2015} suggests a smaller population of CBSSs in the Galaxy than originally thought, but predicts a larger sample when compared to the X-ray-selected CBSSs in a few nearby spiral galaxies \citep{Galiullin2021}. This points to the need to use the other recognized features of these sources to obtain a working sample of CBSSs in our galaxy. The current growth of all-sky spectroscopic optical and \textit{IR} surveys presents an opportunity to search for these sources in large samples by looking for CBSSs' conspicuous spectral features. 

\section{Conclusions}

ZTFJ04 is proposed as a new galactic CBSS candidate. Considering its position near the galactic plane, the current non-detection in X-rays is not a deterrent for such a classification. The object's optical spectrum and SED show a very close resemblance with those seen in well-known CBSSs, especially that of QR\,And. An orbital period of 5.42\,h is derived from a quasi-sinusoidal photometric modulation. Finding these systems by their optical properties may suggest that the galactic population of CBSSs is mostly unseen in X-rays. A follow-up X-ray observation would be ideal for a better understanding of such possible emission from this candidate.

\section*{Acknowledgements}

GRCB acknowledges support from the Conselho Nacional de Desenvolvimento Científico e Tecnológico (CNPq) through the PIBIC programme. MPD acknowledges support from CNPq under grant \#309035. NP and LM acknowledge that this study was financed in part by the Coordenação de Aperfeiçoamento de Pessoal de Nível Superior - Brasil (CAPES) - Finance Code 001. PES, FD and FJ thank the Brazilian Ministry of Science, Technology and Innovation (MCTI) and the Brazilian Space Agency (AEB), who supported the present work under the PO 20VB.0009. The authors thank an anonymous reviewer for her/his helpful comments and suggestions.

Guoshoujing Telescope (the Large Sky Area Multi-Object Fiber Spectroscopic Telescope LAMOST) is a National Major Scientific Project built by the Chinese Academy of Sciences. Funding for the project has been provided by the National Development and Reform Commission. LAMOST is operated and managed by the National Astronomical Observatories, Chinese Academy of Sciences.

This paper includes data collected with the TESS mission, obtained from the MAST data archive at the Space Telescope Science Institute (STScI). Funding for US Institutions for the TESS mission is provided by the NASA Explorer Program. STScI is operated by the Association of Universities for Research in Astronomy, Inc., under NASA contract NAS 5–26555.

Based on observations obtained with the Samuel Oschin Telescope 48-inch and the 60-inch Telescope at the Palomar Observatory as part of the Zwicky Transient Facility project. ZTF is supported by the National Science Foundation under Grants No. AST-1440341 and AST-2034437 and a collaboration including current partners Caltech, IPAC, the Oskar Klein Center at Stockholm University, the University of Maryland, University of California, Berkeley, the University of Wisconsin at Milwaukee, University of Warwick, Ruhr University, Cornell University, Northwestern University and Drexel University. Operations are conducted by COO, IPAC, and UW. 

 This work has made use of data from the European Space Agency (ESA) mission \textit{Gaia} (\url{https://www.cosmos.esa.int/gaia}), processed by the \textit{Gaia} Data Processing and Analysis Consortium (DPAC, \url{https://www.cosmos.esa.int/web/gaia/dpac/consortium}). Funding for the DPAC has been provided by national institutions, in particular the institutions participating in the \textit{Gaia} Multilateral Agreement.

This work has made use of archival data from the \textit{ROSAT}/PSPC All-Sky Survey mission maintained by the Max-Planck-Institut für extraterrestrische Physik (MPE), and hosted by the High Energy Astrophysics Science Archive Research Center (HEASARC), provided by NASA's Goddard Space Flight Center. The \textit{ROSAT} project was supported by the Ministerium für Bildung, Wissenschaft, Forschung und Technologie (BMBF/DARA) and by the Max-Planck-Gesellschaft.  

This research has made use of the NASA/IPAC Infrared Science Archive, which is operated by the Jet Propulsion Laboratory, California Institute of Technology, under contract with the National Aeronautics and Space Administration. 

\section*{Data Availability}

The data underlying this article are public; the references and/or paths to access are provided in the main text.


\bibliographystyle{mnras}
\bibliography{bibliography}

\bsp	
\label{lastpage}
\end{document}